\documentclass[iop]{emulateapj-rtx4}

\usepackage{graphicx}  
\usepackage{dcolumn}   
\usepackage{bm}        
\usepackage{amsfonts,amsmath,amssymb,mathrsfs}
\usepackage{color}

\usepackage{natbib,times}
\definecolor{orange}{rgb}{1,0.5,0}

\newcommand{\fracb}[2]{\left(\frac{#1}{#2}\right)}

\shorttitle{GRBs from Magnetic Reconnection: Variability and Robustness of Lightcurves}
\shortauthors{Granot}

\begin{document}

\title{Gamma-Ray Bursts from Magnetic Reconnection: Variability and Robustness of Lightcurves}

\author{Jonathan Granot}

\affil{Department of Natural Sciences, The Open University of Israel, 1 University Road, 
P.O. Box 808, Ra'anana 4353701, Israel.}

\begin{abstract}
The dissipation mechanism that powers gamma-ray bursts (GRBs) remains uncertain almost half a century after their discovery. 
The two main competing mechanisms are the extensively studied internal shocks and the less studied magnetic reconnection. 
Here we consider GRB emission from magnetic reconnection accounting for the relativistic bulk motions that it produces in 
the jet's bulk rest frame. Far from the source the magnetic field is almost exactly normal to the radial direction, 
suggesting locally quasi-spherical thin reconnection layers between regions of oppositely directed magnetic field.
We show that if the relativistic motions in the jet's frame are confined to such a quasi-spherical uniform layer, 
then the resulting GRB lightcurves are independent of their direction distribution within this layer.
This renders previous results for a delta-function velocity-direction distribution \citep{BG16} applicable to a 
much more general class of reconnection models, which are suggested by numerical simulations. 
Such models that vary in their velocity-direction distribution differ mainly in the size of the bright region 
that contributes most of the observed flux at a given emission radius or observed time. 
The more sharply peaked this distribution, the smaller this bright region, and the stronger the lightcurve variability 
that may be induced by deviations from a uniform emission over the thin reconnection layer, which may be expected 
in a realistic GRB outflow. This is reflected both in the observed image at a given observed time 
and in the observer-frame emissivity map at a given emission radius, which are calculated here for three simple 
velocity-direction distributions. 
\end{abstract}


\maketitle

\section{Introduction}

Gamma-ray Bursts (GRBs) are the most luminous cosmic explosions, with huge isotropic-equivalent luminosities 
of $L_{\rm iso}\sim10^{50}-10^{53}\;{\rm erg\;s^{-1}}$ \citep[for a review see, e.g.,][]{Piran04,KZ15}. 
They divide into two main sub-classes \citep{Kouveliotou93}: 
long-duration ($\gtrsim2\;$s) soft-spectrum GRBs that are associated with broad-lined SNe Ic, implying a
massive-star progenitor \citep[e.g.,][]{WB06}, and short-duration ($\lesssim2\;$s) hard-spectrum GRBs that are thought 
to arise from the merger of a binary neutron-star system or a neutron star and a stellar-mass black hole
\citep{Eichler89,NPP92,LR-R07,Nakar07}. In both classes the central engine is a newly formed rapidly accreting 
stellar-mass black hole or a rapidly rotating highly magnetized neutron star (millisecond magnetar), which launches a 
relativistic jet. 

The bright GRB prompt $\gamma$-ray emission shows rapid variability and typically peaks at photon energies 
of hundreds of keV. This would imply a huge optical depth to pair production, which is incompatible with its non-thermal 
spectrum (the compactness problem), unless the emitting region moves toward us with an ultra-relativistic Lorentz 
factor of $\Gamma\gtrsim100$ \citep{BH97,LS01,Granot08b,Hascoet12}. Such a highly relativistic outflow also 
naturally explains the subsequent afterglow emission in X-ray, optical and radio over days, weeks and months after the GRB,
as the ejecta are decelerated by the external medium and drive a long-lived shock into it, 
which gradually decelerates as it sweeps-up more mass. Compactness arguments also require
a large enough prompt emission radius ($\sim10^{13}-10^{16}\;$cm) in particular for the $\sim\;$GeV photons detected 
by \textit{Fermi} in some GRBs \citep[e.g.,][]{GRB080825C,GRB081024B,LAT_catalog}.  The observed fast variability 
of the GRB prompt emission implies that it must be produced by internal dissipation within the ejecta \citep{SP97}. 

The GRB outflow composition, as well as the dissipation and emission mechanisms that produce the prompt emission
are still uncertain, and are important open questions in this field. They can also affect each other, as the outflow composition 
affects its dynamics and dissipation, which in turn affect the resulting emission. In particular, a key question is whether the 
energy is carried out from the central source to the emission region predominantly as kinetic energy -- a baryonic jet 
\citep{Goodman86,Paczynski86,SP90}, or as Poynting flux -- a highly magnetized (or Poynting-flux dominated) jet \citep{Usov92,Thompson94,MR97,Blandford02,Lyutikov06,Granot15} with a large magnetization parameter $\sigma$ 
(the magnetic-to-particle enthalpy density or energy flux ratio). A baryonic, kinetically dominated jet can naturally 
lead to reasonably efficient energy dissipation via internal shocks within the outflow \citep{RM94}. This may also occur 
in an initially high-$\sigma$ outflow that is highly variably, due to impulsive acceleration that converts its initial 
magnetic energy into kinetic energy \citep{GKS11,Granot12}. As long as the flow remains highly magnetized this suppresses 
internal shocks. On the other hand, in high-$\sigma$ outflows there is an alternative dissipation
mechanism that can be more efficient than internal shocks -- magnetic reconnection \citep{Thompson94,Spruit01,LB03,GS07,Lyubarsky10,Kagan15}.

A high $\sigma$ near the central source can help avoid excessive baryon loading that might prevent the jet from reaching
sufficiently high Lorentz factors far from the source, at the emission region. Such initially high-$\sigma$ jets are also favored 
on energetic grounds, since modeling of GRB central engines that rely on hydromagnetic jet launching via accretion disks 
suggest that their power is significantly larger than that of thermally driven outflows powered by neutrino-anti neutrino 
annihilation \citep[e.g.,][]{KPK13}, and they may naturally lead to magnetic reconnection. 

In a striped wind magnetic field configuration \citep[e.g.][]{Coroniti90}, whether the flipping of the magnetic field direction 
near the source is periodic (as expected for a millisecond-magnetar central engine) or stochastic (as expected for an 
accreting black hole), reconnection at large distances from the source has a natural preferred direction. At such large 
distances the magnetic field is almost exactly normal to the (spherical) radial direction, as are the current sheets that 
separate regions of opposite magnetic polarity where reconnection occurs, thus forming nearly spherical thin
reconnection layers. Moreover, for a large $\sigma$ just before the dissipation region reconnection leads to local
relativistic bulk motion of the outgoing particles away from the reconnection sites in the jet's bulk frame, with a Lorentz 
factor $\Gamma'$ that can reach a few to several. This leads to anisotropic emission in the jet's bulk frame, which can
significantly affect the observed emission. 

\begin{figure}
\begin{center}
\includegraphics[angle=0,width=0.48\textwidth]{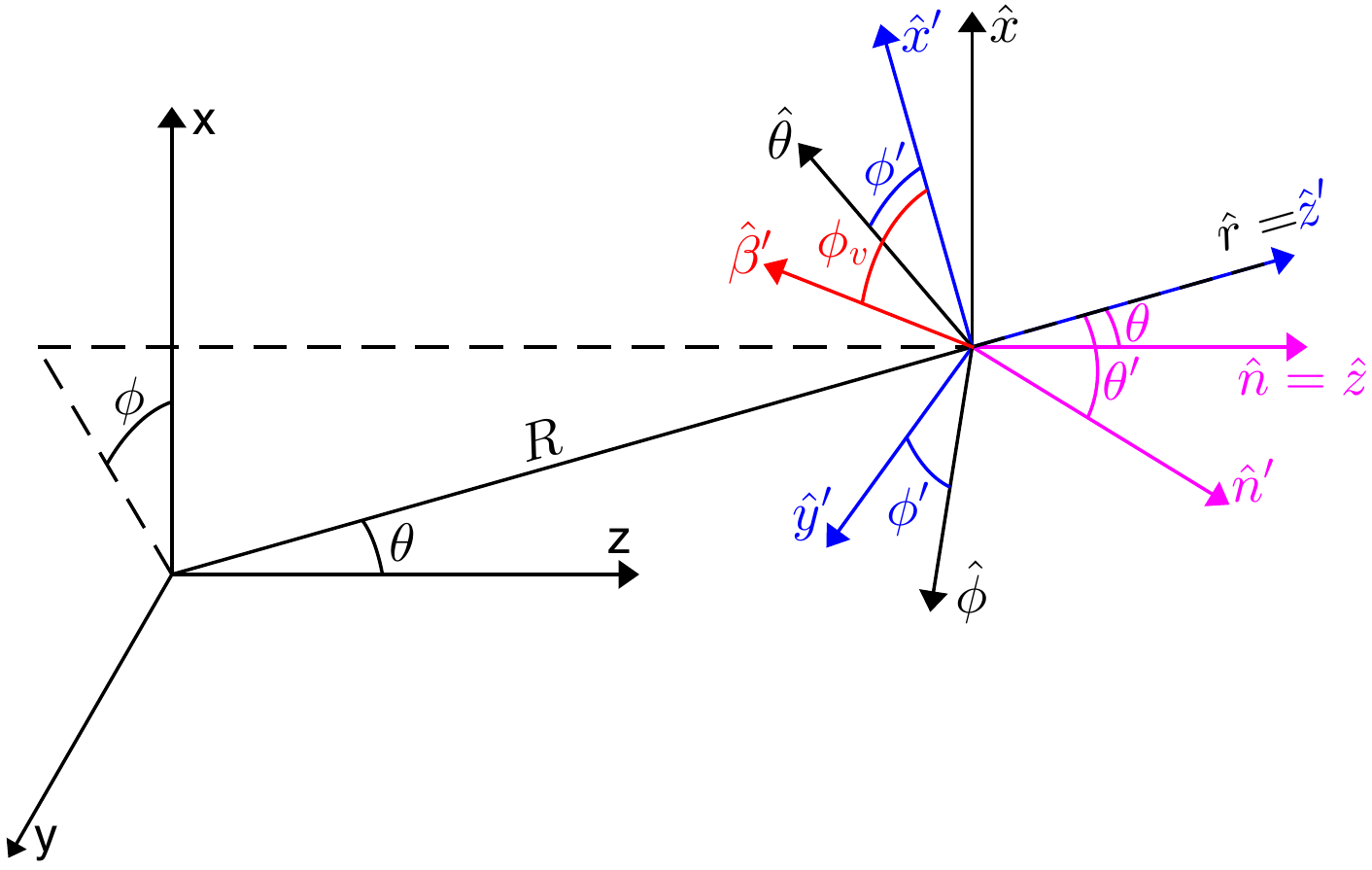}
\includegraphics[angle=0,width=0.48\textwidth]{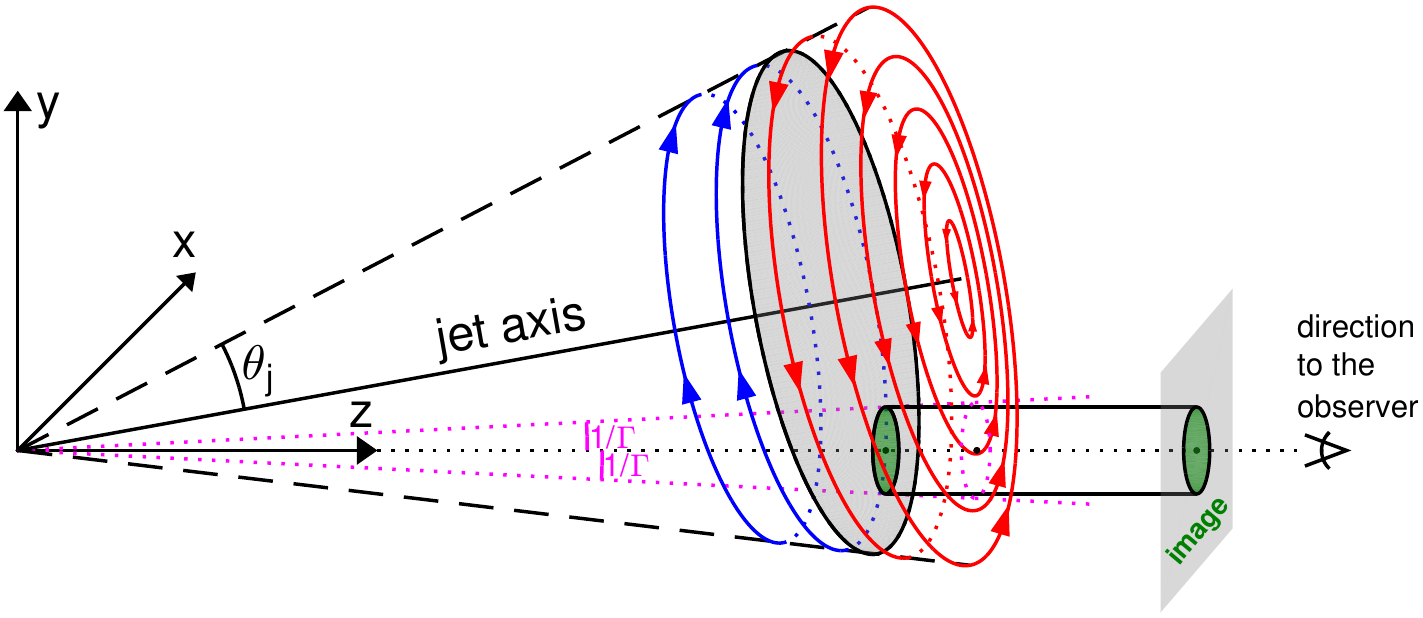}
\end{center}\caption{Schematic geometry of our model. {\bf{}Bottom}: 
Our basic model features shells of oppositely oriented toroidal magnetic field (in \textit{blue} and \textit{red}),
separated by a locally quasi-spherical thin reconnection layer (in \textit{gray}). The observed region of angle $\sim1/\Gamma$
around the line of sight (in \textit{green}) that contributes to the observed image and flux is a small part of the jet.
{\bf{}Top}: the unprimed source rest frame (in \textit{black}) 
is shown in Cartesian $(x,y,z)$ and spherical $(R,\theta,\phi)$ coordinates, where $\hat{z}$ points to the observer.
The primed, jet's bulk frame (in \textit{blue}) is the local rest frame of a point on a spherical emitting shell of radius $R$ 
expanding radially with Lorentz factor $\Gamma\gg1$. The velocity direction $\hat{\beta}'$ (in \textit{red}) of the 
emitting plasma in the primed frame is in the $x'$--$y'$ plane (normal to the radial direction) at an angle $\phi_v$ 
from $\hat{x}'$ (the local magnetic-field direction before reconnection). 
Also shown (in \textit{magenta}) are the directions of a photon that reaches the observer in both frames
(which are related through aberration of light).} 
\label{fig:geom}
\vspace{0.5cm}
\end{figure}

Figure\;\ref{fig:geom} (bottom panel) shows a simple manifestation of our basic model where the jet consists of 
shells with oppositely oriented toroidal magnetic field, separated by quasi-spherical current sheets where reconnection occurs 
(a modest poloidal field-component should not significantly change this basic picture). In GRBs the jet half-opening angle 
typically satisfies $\theta_j\gg1/\Gamma$ so only a small fraction of the jet ($\sim(\Gamma\theta_j)^{-2}\ll1$; the green circle
in Figure\;\ref{fig:geom}) is visible, and the magnetic field may be approximated as uniform within it. This approximation was made 
for calculating the prompt-GRB polarization \citep{Granot03,GK03}, and should not greatly affect our results. For the afterglow 
polarization the global toroidal-field structure was considered \citep{Lazzati04,GT05} since the whole jet becomes visible as it 
decelerates during the afterglow. Anisotropic synchrotron emission was considered as a possible cause of early X-ray afterglow  
variability or rapid decay \citep{Beloborodov11}. We allow for any reconnection-induced velocity-direction distribution in the jet's 
bulk frame $g(\phi_v)$ within the quasi-spherical reconnection layer ($\phi_v$ is defined in Figure\;\ref{fig:geom}, top panel).
Such an anisotropic emission model was recently considered for the prompt-GRB emission by \citealt{BG16} (hereafter BG16), where 
velocities are in the direction of the anti-parallel magnetic-field lines just prior to their reconnection, which is uniform within visible region.

Our anisotropic emission model differs from previous relativistic-turbulence models \citep{LB03,KN09,LNP09} 
that assume an isotropic velocity distribution of the motions in the jet's bulk frame.
For this model BG16 calculated the expected lightcurves and spectra of the prompt emission, and demonstrated  
that it can potentially reproduce many of the observed prompt GRB properties (e.g. its variability, pulse asymmetry, the 
very rapid decay phase at its end, and many of the observed correlations).

Recent simulations of relativistic magnetic reconnection suggest that as $\sigma$ increases, both the reconnection
rate and resulting particle bulk velocities ($\beta'$) increase, and the power-law index of their energy spectrum becomes 
harder \citep{Cerutti12,Cerutti14,Sironi14,Guo15,Kagan15,Liu15}. In high-$\sigma$ GRB outflows one may typically 
expect $\Gamma'\sim\;$a few to several. The collimation of the accelerated electrons appears to increase with their energy. 
Their velocities are indeed predominantly confined to the reconnection layer, but are not necessarily along the anti-parallel
directions of the magnetic field lines just before the reconnection (as was assumed by BG16). 
This motivates us to consider such velocity distributions that are more general.

In Section~\ref{sec:LCindependence} the lightcurve is shown to be independent of the angular distribution 
$g(\phi_v)$ of the velocities in the jet's bulk frame as long as they are confined to a uniformly emitting 
spherical reconnection layer; $g(\phi_v)$ does, however, affect the observed image and the contribution to the observed 
flux from a given emission radius, which are calculated in Sections~\ref{sec:image} and \ref{sec:FR}, respectively. 
This may in turn affect the prompt GRB lightcurve if the emission across the spherical
reconnection layer is non-uniform, which may be expected under realistic conditions. 
Finally, the main results are summarized and discussed in Section~\ref{sec:dis}.

\section{Flux Density is Independent of Velocity Directions within a Uniform Spherical Reconnection Layer}
\label{sec:LCindependence}

Here we show that the observed flux density $F_\nu(T)$ at any observed frequency $\nu$ and time $T$ 
is independent of the velocity-direction distribution of the emitting plasma within a uniform spherical thin 
reconnection layer. Let $g(\phi_v)$ be such a general probability distribution  (normalized as 
$\int_0^{2\pi}g(\phi_v)d\phi_v=1$) of local velocity directions in the jet's bulk frame 
(that is primed in Figure\;\ref{fig:geom}, top panel) that are at angles $\phi_v$ 
relative to the local direction of the magnetic field ($\hat{x}'$ in Figure\;\ref{fig:geom}, which is a preferred 
direction within the reconnection layer, and is assumed here to be uniform within the visible region). 
We follow the notations of BG16 (e.g., in the source's frame $\theta$ is the polar angle 
measured from the line of sight, and $\phi$ is the azimuthal angle). The general expression for the flux 
density is then given by a weighted average over that for a single velocity direction taken from BG16,
\begin{eqnarray}\nonumber
F_\nu(T)=&\frac{L''_{\nu''_0}}{(4\pi{}D)^2}\int{}dy\left|\frac{d\mu}{dy}\right|\mathcal{D}^3(y)f\left[y\fracb{T}{T_0}^\frac{1}{m+1}\right]
\\
&\quad\times\int_0^{2\pi}d\phi\int_0^{2\pi}d\phi_v\,g(\phi_v)S(x)\frac{\mathcal{D}'^{\,3-k}}{\Gamma'^{\,k}}\;,
\end{eqnarray}
where $D=d_L(1+z)^{-1/2}$ is the effective distance to the source and $d_L(z)$ is the luminosity distance, 
$y=R/R_L$ is the normalized radius, $k=0$ for a blob and $k=1$ for a steady state in the jet's frame, $\mu=\cos\theta$, 
$\mathcal{D}(y)=1/\Gamma(1-\beta\mu)$ is the Doppler factor between the rest frame of the central 
source and the jet's bulk frame, $\mathcal{D}'=1/\Gamma'[1-\beta'\sin\theta'\cos(\phi-\phi_v)]$ is the 
Doppler factor between the jet's bulk frame and the local emitting plasma's rest frame (it depends on $y$ through $\theta'$), 
and $x=\nu''/\nu''_0(y)=\nu_z/[\mathcal{D}\mathcal{D}'\nu''_0(y)]$ where $\nu_z=(1+z)\nu$ is the frequency
in the source's cosmological frame. 
Thus, the only dependence on the azimuthal angle $\phi$ is through $\mathcal{D}'$, both directly and through $x$,
and this dependence is in turn only through $\varphi\equiv\phi-\phi_v$.
Therefore, one can reverse the order of integration over $\phi$ and $\phi_v$, and change variables from $\phi$ to $\varphi$,
\begin{equation}
\int_0^{2\pi}d\phi_vg(\phi_v)\,\int_0^{2\pi}d\varphi\,S[x(\varphi)]\frac{\mathcal{D}'^{\,3-k}(\varphi)}{\Gamma'^{\,k}}\;,
\end{equation}
where the inner integral over $\varphi$ is independent of $\phi_v$, so that the outer integral over $\phi_v$ gives 1
from the normalization of $g(\phi_v)$. This reduces the expression for the observed flux density
to that for a delta function in velocity direction  (e.g. $g_1(\phi_v)$ in Eq.~[\ref{eq:g}]) as in BG16,
where one can take $\phi_v=0$,
\begin{equation}\label{eq:single_v}
F_\nu(T)=\frac{L''_{\nu''_0}}{(4\pi D)^2}\int{}dy\left|\frac{d\mu}{dy}\right|\mathcal{D}^3f(y)\int_0^{2\pi}d\phi S(x)\frac{\mathcal{D}'^{\,3-k}}{\Gamma'^{\,k}}\;.
\end{equation}
The reason why the observed flux is independent of $g(\phi_v)$ is as follows. The observed flux is the weighted mean 
of the contributions from plasma with different velocity directions $\phi_v$. However,  the observed flux from such a 
uni-directional distribution does not depend on its absolute direction $\phi_v$, since the latter affects only the dependence 
of the observed radiation on the azimuthal angle $\phi$, and thus the observed image, but not the photon arrival times or the
observed flux density.

\section{The Observed Image for Anisotropic Emission}
\label{sec:image}

This motivates us to calculate the observed image for different choices of $g(\phi_v)$ and $\Gamma'$.
For comparison we will also show the image for isotropic emission in the jet's bulk frame ($\Gamma'=1$), from 
\cite{Granot08}. In particular, we will use
\begin{eqnarray}\nonumber
g_1(\phi_v)=\frac{\delta(\phi_v)+\delta(\phi_v-\pi)}{2}\ ,\quad\quad\ 
\\ \label{eq:g}
g_2(\phi_v)=\frac{\cos^2\phi_v}{\pi}\ ,\quad g_3(\phi_v)=\frac{1}{2\pi}\;,
\end{eqnarray}
where $g_1(\phi_v)$ (used in BG16) corresponds to velocity along the anti-parallel reconnecting magnetic field lines,
$g_2(\phi_v)$ is motivated by PIC simulations of relativistic reconnection, and $g_3(\phi_v)$ is the extreme assumption 
of a uniform velocity distribution within the thin reconnection layer. For each of these $g(\phi_v)$
we calculate the image for $\Gamma'=1,\,2,\,4,\,8$. 

\begin{figure}
\begin{center}
\includegraphics[angle=0,width=0.48\textwidth]{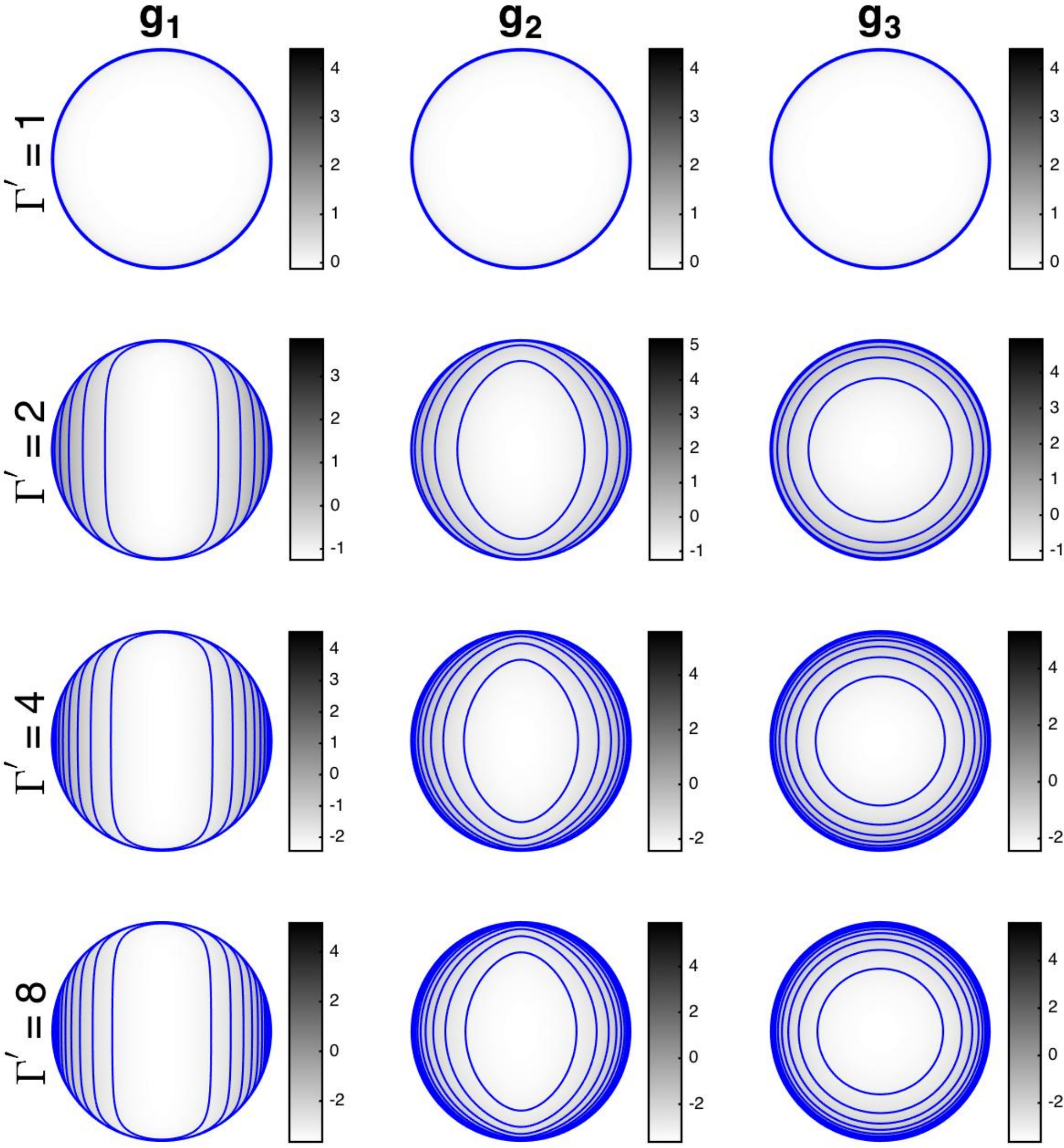}
\end{center} \caption{Images (according to Equation~(\ref{eq:Inu}), 
corresponding to the green region in Figure\;\ref{fig:geom}, bottom panel) 
for different velocity distributions within the reconnection layer (functions $g_{1}$, $g_{2}$ 
and $g_{3}$ in Eq.$\;$(\ref{eq:g}), from left to right), and for different values of 
$\Gamma'$ ($=1,\,2,\,4,\,8$ from top to bottom), for $m=0$ ($\Gamma\propto{}R^{-m/2}$), $k=1$ (steady-state 
reconnection in the jet's bulk frame), $a=0$ and $\alpha=1$ ($L''_{\nu''}\propto{}R^a(\nu'')^{-\alpha}$). 
Shown are logarithmic maps of the specific intensity normalized by its mean value in the image,
$\log_{10}(I_\nu/\langle{}I_\nu\rangle)$, with contours at $\log_{10}[I_\nu/\min(I_\nu)]=0.5,1,1.5,...$.} 
\label{fig:fig1}
\vspace{0.5cm}
\end{figure}

The flux density differential is $dF_\nu=I_\nu{}d\Omega=I_\nu{}dS_\perp/d_A^2$, where $d_A(z)$ is the
angular distance to the source and $S_\perp$ is the area of the image, normal to the line of sight. If $R_\perp$ is the 
corresponding distance from the center of the image, then 
\begin{equation}
dS_\perp=R_\perp{}dR_\perp{}d\phi=\fracb{R_L}{\Gamma_L}^2\frac{\left[1-(m+2)y^{m+1}\right]}{2(m+1)}dyd\phi\;.
\end{equation}
where $\Gamma^2\propto{}R^{-m}$. 
We are interested in the specific intensity at a general location within the image, $I_\nu(r,\phi)$, where 
\begin{equation}\label{eq:r(y)}
r\equiv\frac{R_\perp}{R_{\perp,\rm{max}}}=\frac{(m+2)^{\frac{m+2}{2(m+1)}}}{\sqrt{m+1}}\sqrt{y-y^{m+2}}\;,
\end{equation}
and $R_{\perp,\rm{max}}=(m+2)^{-(m+2)/[2(m+1)]}R_L/\Gamma_L$. As we evaluate $I_\nu$ at a fixed $\phi$,
one still needs to integrate over $\phi_v$, or more conveniently switch variables to $\varphi$ and obtain
\begin{equation}
\frac{dF_\nu}{dyd\phi}=\frac{L''_{\nu''_0}}{(4\pi{}D)^2}\left|\frac{d\mu}{dy}\right|\mathcal{D}^3f(y)
\int_0^{2\pi}d\varphi\,\frac{g(\phi-\varphi)S[x(\varphi)]}{\Gamma'^{\,k}\mathcal{D}'^{\,k-3}(\varphi)}\;,
\end{equation}
\begin{eqnarray}\nonumber
I_\nu=&\frac{L''_{\nu''_0}}{(4\pi)^2}\frac{\mathcal{D}^3f(y)}{(1+z)^3}\left|\frac{d\mu}{dy}\right|\fracb{\Gamma_L}{R_L}^2\frac{2(m+1)}{|1-(m+2)y^{m+1}|}
\\ 
&\quad\times\int_0^{2\pi}d\varphi\,\frac{g(\phi-\varphi)S[x(\varphi)]}{\Gamma'^{\,k}\mathcal{D}'^{\,k-3}(\varphi)}\;,
\end{eqnarray}
Now we shall use the expressions for the relevant terms,
\begin{equation}
\left|\frac{d\mu}{dy}\right|=\frac{y^{-2}+my^{m-1}}{2(m+1)\Gamma_L^2}\;,\quad
\mathcal{D}=\Gamma_L\frac{2(m+1)y^{-m/2}}{m+y^{-m-1}}\;,
\end{equation}
\begin{equation}
\mathcal{D}'(\varphi)=\frac{1}{\Gamma'}\left(1-2\beta'\frac{\sqrt{(m+1)(y^{-m-1}-1)}}{m+y^{-m-1}}\cos\varphi\right)^{-1}\;.
\end{equation}
Now, for simplicity, we shall specify to a power-law spectrum, $S(x)=x^{-\alpha}$, and emission with radius,
$f(R/R_0)\propto{}R^a$ between $R_0$ and $R_f=R_0+\Delta{}R$, with a constant $\Gamma'$ and 
$\nu''_0$,~\footnote{This result reduces to Eq.\;(15) of \cite{Granot08} for isotropic emission in the jet's bulk frame 
($\Gamma'=1$), with the small modifications given in Eqs.\;(8) and (17) therein, which reflect the difference between 
a shock and a reconnection layer. To match the notations there one should take $\alpha\to-b$ and $m\to3-k$ where there 
$k$ is the power-law index of the external density profile in front of the afterglow shock.}
\begin{equation}\label{eq:Inu}
I_\nu\propto\frac{\nu^{-\alpha}\,T^\frac{2a-4-m(3+\alpha)}{2(m+1)}y^{a-1-\frac{m}{2}(1+\alpha)}}
{(m+y^{-m-1})^{2+\alpha}\left|1-\frac{m+2}{y^{-m-1}}\right|}\int_0^{2\pi}\frac{g(\phi-\varphi)d\varphi}{\mathcal{D}'^{\,k-3\alpha(\varphi)}}\;.
\end{equation}

\begin{figure}
\begin{center}
\includegraphics[angle=0,width=0.48\textwidth]{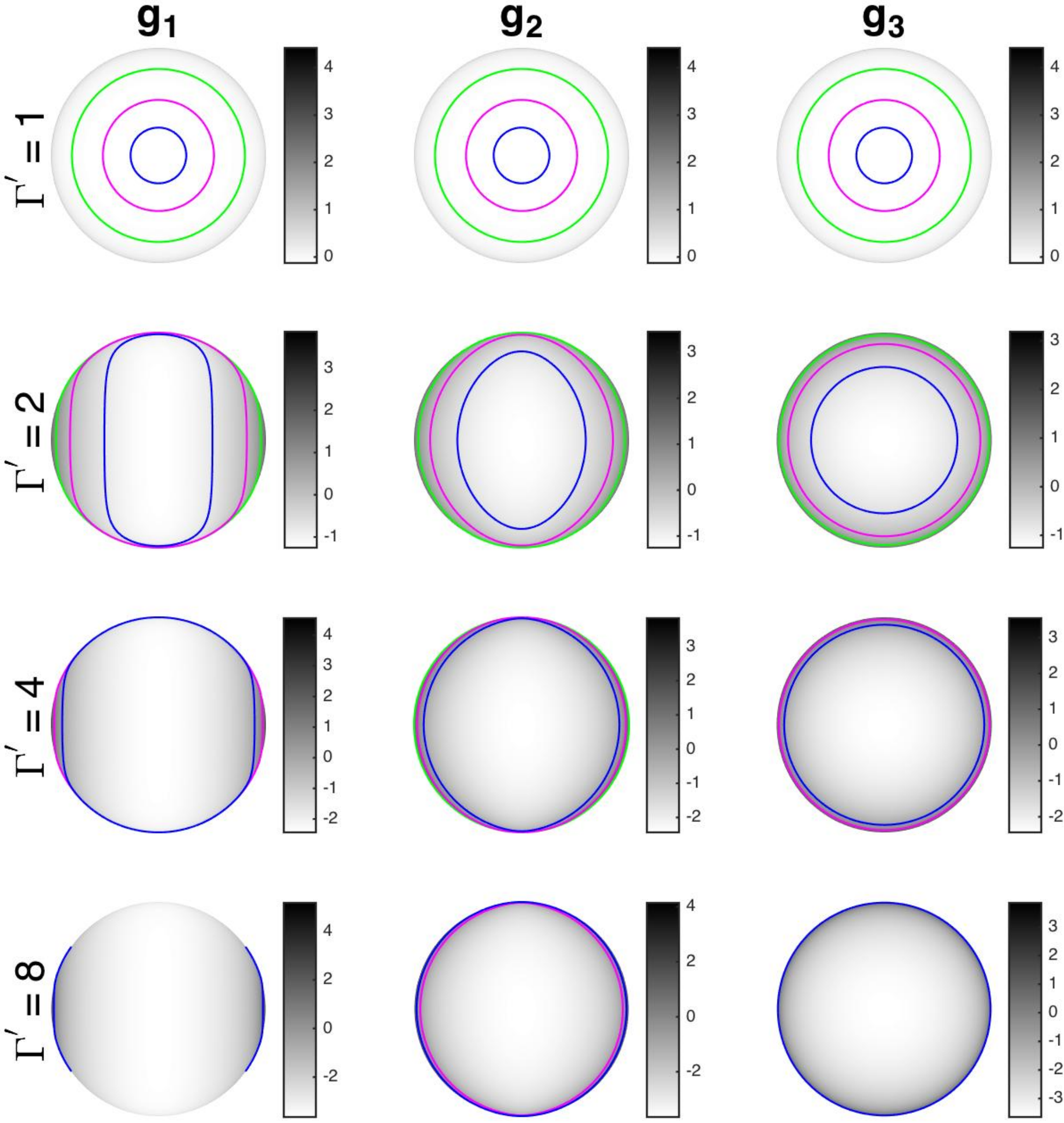}
\end{center} \caption{Similar to Figure\;\ref{fig:fig1}, but showing contour lines for which 50\% (\textit{green}), 
80\% (\textit{magenta}), and 95\% (\textit{blue}) of the total flux comes from higher $I_\nu$ values, i.e. from the 
region between the contour and the outer edge of the image.} 
\label{fig:fig2}
\vspace{0.5cm}
\end{figure}

Each $r<1$ corresponds to two values of $y$, at the front ($y_{+}$) and the back ($y_{-}$) of the equal arrival time 
surface of photons to the observer. They are generally found by numerically solving Eq.\;(\ref{eq:r(y)}), but for 
some $m$-values $y_\pm(r;m)$ can be found analytically \citep{Granot08}, e.g. $y_\pm(r;0)=\frac{1}{2}(1\pm\sqrt{1-r^2})$
and $y_\pm(r;1)=(2/\sqrt{3})\cos\left[\frac{1}{3}\left(\pi\mp\arctan\sqrt{r^{-4}-1}\,\right)\right]$. One must add up 
these two contributions to $I_\nu(r,\phi)$. There is contribution only from radii $R_0\leq{}R\leq{}R_f$ corresponding to
$y_{\rm min}\leq{}y\leq{}y_{\rm max}$ where $y_{\rm min}=\min[1,R_0/R_L(T)]$ and $y_{\rm max}=\min[1,R_f/R_L(T)]$.
In the following, for simplicity, emission is assumed from all radii.
 
The resulting images are shown in Figs.\;\ref{fig:fig1} and \ref{fig:fig2}.
Figure\;\ref{fig:fig1} adds equally spaced contour lines, with $\Delta\log_{10}(I_\nu)=0.5$. 
Figure\;\ref{fig:fig4} adds contour lines at $I_\nu$ values above which 50\% ({\it green}), 
80\% ({\it magenta}), and 95\% ({\it blue}) of the total flux originates.
For $\Gamma'\gtrsim2$ most of the flux clearly comes from a small part of the image near its outer edge.
For $g_1(\phi_v)$ (a delta-function anti-parallel velocity distribution) most of the flux comes from 
two small regions near the outer edge of the image, which quickly decrease in size as $\Gamma'$ increases.
For $g_2(\phi_v)=\frac{1}{\pi}\cos^2\phi_v$ most of the flux comes from an asymmetric ring at the outer
edge of the image. For $g_3(\phi_v)=1/2\pi$ (an isotropic velocity distribution within the reconnection layer 
-- normal to the radial direction) this ring becomes symmetric about the center of the image,
following the behavior of the whole image in this case for which there is no preferred $\phi$-direction.

\begin{figure}
\begin{center}
\includegraphics[angle=0,width=0.48\textwidth]{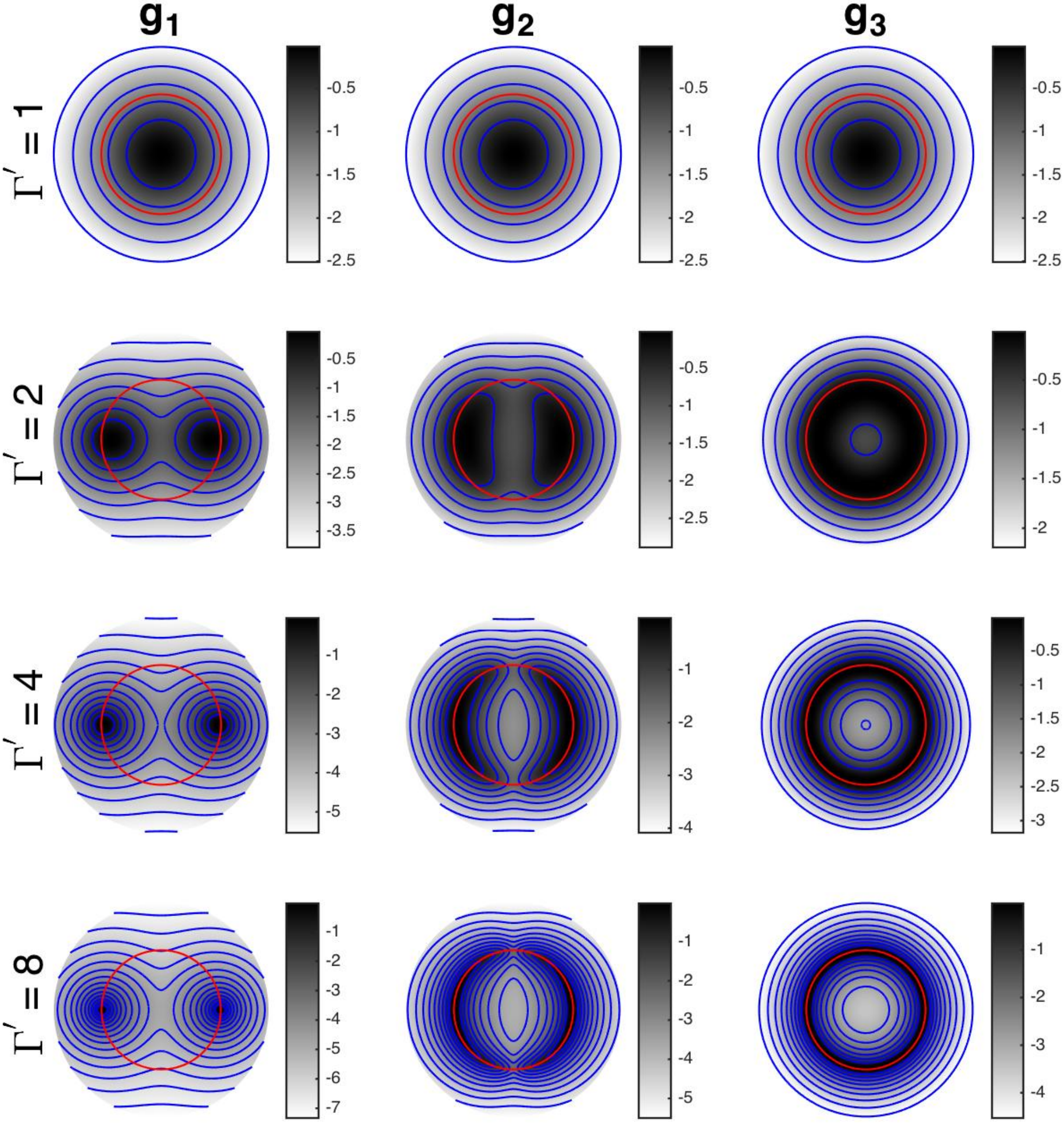}
\end{center} \caption{Logarithmic maps of the (normalized) contribution to the observed flux density per unit area of 
the shell at a given radius $R$, $\log_{10}[(dF_\nu/dA)/\max(F_\nu/dA)]$ according to Eq.\;(\ref{eq:constR}), 
for different velocity distributions within the reconnection layer (functions $g_{1}$, $g_{2}$ and $g_{3}$ in Eq.\;(\ref{eq:g}), 
from left to right), and for different values of $\Gamma'$ ($=1,\,2,\,4,\,8$ from top to bottom), for $m=0$, $k=1$, $a=0$ 
and $\alpha=1$. The contour lines are at $-\log_{10}[(dF_\nu/dA)/\max(F_\nu/dA)]=0.5,1,1.5,...$. A red circle is added at 
$\theta=1/\Gamma(R)$ for reference.} 
\label{fig:fig3}
\vspace{0.5cm}
\end{figure}

\section{Contribution to Observed Flux from a Given Radius}
\label{sec:FR}

\begin{figure}
\begin{center}
\includegraphics[angle=0,width=0.48\textwidth]{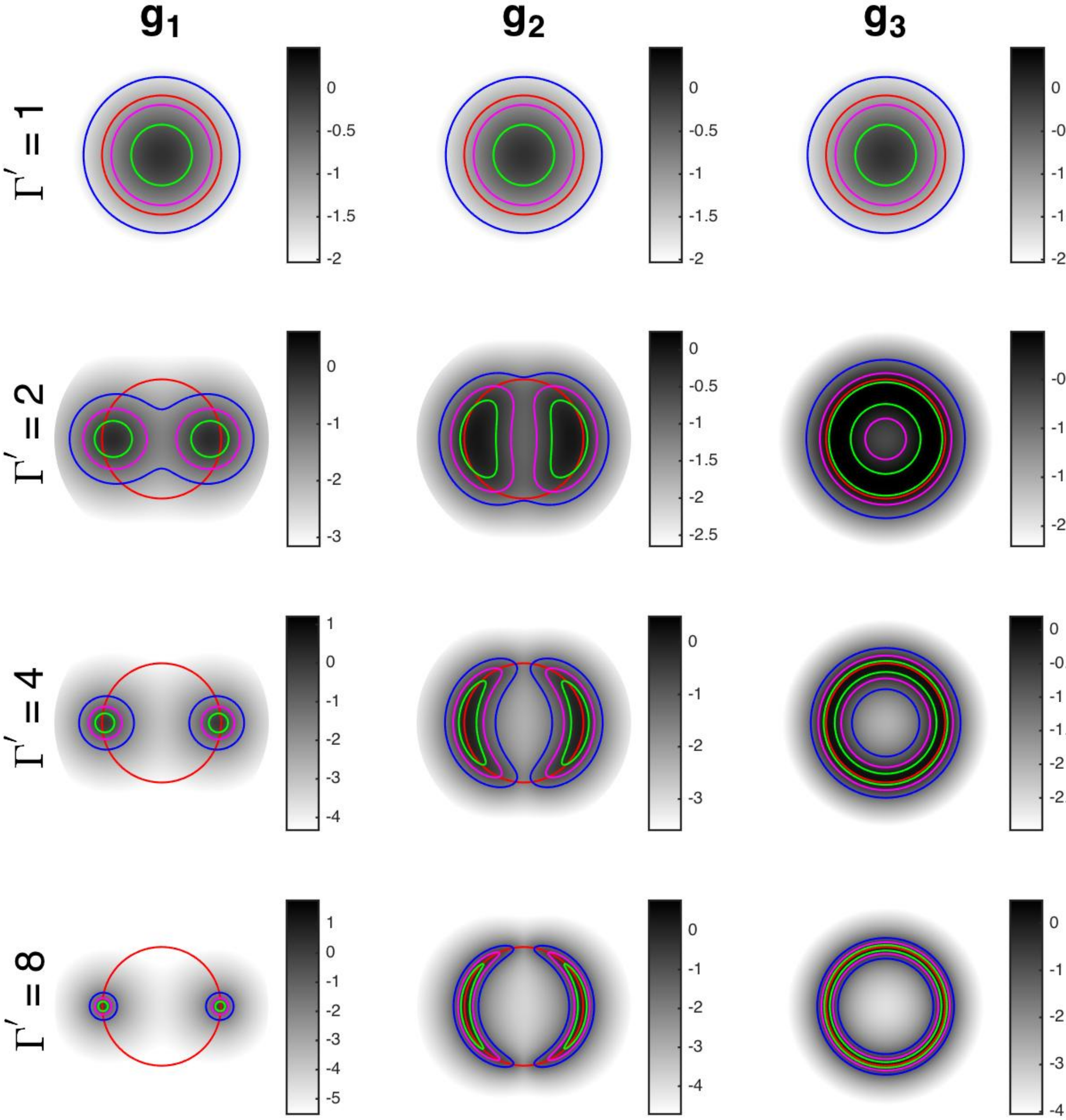}
\end{center}\caption{Similar to Figure\;\ref{fig:fig3}, but showing contour lines at the $dF_\nu/dA$ values above which 
50\% ({\it green}), 80\% ({\it magenta}), and 95\% ({\it blue}) of the total flux from the given emission radius originates,
corresponding to the regions enclosed by these contours.} 
\label{fig:fig4}
\vspace{0.5cm}
\end{figure}

It is also useful to examine the contribution to the observed flux from a given emission radius $R$ 
(as a function of $\theta$ and $\phi$) even though it arrives
over a range of observed times $T$. To this end we consider the contribution per unit area of the shell 
$dA=R^2d\mu d\phi$ at a constant $R$ and $\Gamma=\Gamma(R)$, where
\begin{equation}
\mathcal{D}=\frac{2\Gamma}{1+(\Gamma\theta)^2}\ ,\quad
\mathcal{D}'=\frac{1}{\Gamma'\left(1-\beta'\frac{2(\Gamma\theta)}{1+(\Gamma\theta)^2}\cos\phi\right)}\;.
\end{equation}
Altogether, for a power-law emission spectrum one obtains
\begin{equation}\label{eq:constR}
\frac{dF_\nu}{d\mu d\phi}\propto\frac{\nu^{-\alpha}}{[1+(\Gamma\theta)^2]^{3+\alpha}}\int_0^{2\pi}\frac{g(\phi-\varphi)d\varphi}
{\left(1-\frac{2(\Gamma\theta)\beta'}{1+(\Gamma\theta)^2}\cos\varphi\right)^{3+\alpha-k}}\;.
\end{equation}
The integrals over $\varphi$ in Eqs.\;(\ref{eq:Inu}) and (\ref{eq:constR}), $G_j(\phi)$ for $g_j(\phi-\varphi)$,
generally give hypergeometric functions for $j = 2,\, 3$. However, for integer $\alpha-k$ values they become particularly 
simple. E.g., for $\alpha-k=0$, $G_3(\phi)\propto(1-B)^{-5/2}(2+B)$ and 
$G_2(\phi)\propto(1-B)^{-5/2}(2+B[1+3\cos(2\phi)])$, where $B=(\beta'\sin\theta')^2$ is given by
\begin{equation}
B=\left\{\begin{matrix}\left(\frac{2\beta'}{m+y^{-m-1}}\right)^2(m+1)(y^{-m-1}-1)&\quad({\rm Eq.}\;(\ref{eq:Inu}))\;,\cr\cr
4(\beta')^2(\Gamma\theta)^2\left[1+(\Gamma\theta)^2\right]^{-2}&\quad ({\rm Eq.}~(\ref{eq:constR}))\;.\end{matrix}\right.
\end{equation}

Figures\;\ref{fig:fig3} and \ref{fig:fig4} show logarithmic color maps of $dF_\nu/dA$,
the contribution to the observed flux density per unit area of the emitting shell from a given radius, 
according to Eq.\;(\ref{eq:constR}).  Figure\;\ref{fig:fig3} adds equally spaced contour lines, with 
$\Delta\log_{10}(dF_\nu/dA)=0.5$.
Figure\;\ref{fig:fig4} adds contour lines at the $dF_\nu/dA$ values above which 50\% ({\it green}), 
80\% ({\it magenta}), and 95\% ({\it blue}) of the total flux from the given emission radius originates.

\section{Discussion}
\label{sec:dis}

In Section~\ref{sec:LCindependence} it was shown that the observed flux density (and thus the lightcurves and spectra)
of GRB prompt emission from a uniform spherical thin reconnection layer are independent of the distribution of velocity
$\vec{\beta}'$ directions within this layer in the jet's bulk frame. This implies that the detailed results  for the 
lightcurves, spectra, and temporal-spectral correlations of BG16, who assumed velocities along two anti-parallel directions, 
are valid for a much larger class of reconnection models, which is consistent with the results of recent simulations.

In Sections~\ref{sec:image} and \ref{sec:FR} it was shown that as $\Gamma'$ increases, the size of the ``bright part" within
the observed region of the reconnection layer that contributes most of the observed flux becomes significantly smaller. 
Moreover, its area and angular size depend on the spread of $\hat{\beta}'$, as expressed in the angular distribution $g(\phi_v)$. 
For $\Gamma'\gtrsim\;$a few, for the tightest angular distribution we considered of two anti-parallel directions ($g_1$ in
Equation\;(\ref{eq:g})) most of the observed flux comes from two small circular regions of angular size $\sim1/(\Gamma'\Gamma)$
(see left panels of Figure\;\ref{fig:fig4}), which occupy a fraction $\sim\Gamma'^{-2}$ of the visible region. 
On the other extreme, for our most spread-out velocity distribution that is uniform within the reconnection layer 
($g_3$ in Equation\;(\ref{eq:g})), most of the flux comes from a thin ring of angular radius $1/\Gamma$ and width 
$\sim1/(\Gamma'\Gamma)$ (see right panels of Figure\;\ref{fig:fig4}), occupying a fraction $\sim1/\Gamma'$ of the visible region. 
These results should not significantly change when relaxing our approximation of a uniform magnetic field within the visible region.

These results may be important if the emission over the spherical thin reconnection layer is not uniform but has some 
angular dependence, e.g. due to irregularities or non-uniformity in the reconnection rate. The value of $\sigma$
affects $\Gamma'$ (which determines the size of the region contributing most of the observed flux), the reconnection rate
(which affects the local radiated power per unity area in the reconnection layer), as well as the electron energy distribution that affects the emission spectrum (and hence the observed spectrum and flux at a given observed energy range). Since $\sigma$ 
may vary with the angular location within the outflow, or even with time at a fixed angular location, one might expect that 
this could potentially lead to significant angular inhomogeneities in the emission from a given radius, as well as temporal changes
at a given angular location. 

If the prompt emission occurs when the jet is coasting at a constant $\Gamma$ then the angular
location of the ``bright part" (which is at an angle of $1/\Gamma$ from the line of sight)
is fixed in time and the lightcurve variability reflects mainly the radial profile of the emission within this small region.
If, on the other hand, the jet is still accelerating or conversely starting to decelerate during the reconnection, then
the ``bright part" will scan through different angular locations and the lightcurve variability could also reflect the
angular distribution of the spectral emissivity in the reconnection layer. In all cases, the larger this ``bright part" 
(i.e. the smaller $\Gamma'$ or $\sigma$, and the wider the velocity spread $g(\phi_v)$)
the more it might average out over different local fluctuations  or angular inhomogeneities in the emission, thus 
reducing the lightcurve variability. Conversely, a larger lightcurve variability may be expected for a smaller
``bright part" (i.e. a larger $\Gamma'$ or $\sigma$, and a narrower velocity spread $g(\phi_v)$), due to
less averaging out, and a larger sensitivity to fluctuations in the emission over small times or angular scales.
A more detailed and quantitative study of these effects on the observed prompt GRB emission is planned in a 
future work.

\acknowledgements 
J.~G. thanks Paz Beniamini for useful comments, and acknowledges support from the ISF grant 719/14.
\\ \vspace{1.0cm}\\

\end{document}